\documentclass[reprint, amsmath, amssymb, aps, prl, superscriptaddress, floatfix, nofootinbib, citeautoscript, longbibliography]{revtex4-1}
\usepackage[utf8]{inputenc}
\usepackage[T1]{fontenc}
\usepackage[parfill]{parskip}
\usepackage{siunitx}
\usepackage{blindtext}
\usepackage[colorlinks, linkcolor=blue, citecolor=blue, urlcolor=blue, breaklinks=true]{hyperref}
\usepackage{graphicx}
\usepackage{dcolumn}
\usepackage{bm}
\usepackage{color}
\usepackage{units}
\usepackage{dsfont}
\usepackage{hyperref}
\usepackage{xcolor}

\newcommand{\bra}[1]{\left\langle #1 \right|}
\newcommand{\ket}[1]{\left| #1 \right\rangle}

\begin{document}

\title{High-dimensional two-photon interference effects in spatial modes}

\author{Markus Hiekkam\"aki}
\email{markus.hiekkamaki@tuni.fi}
\affiliation{Tampere University, Photonics Laboratory, Physics Unit, Tampere, FI-33720, Finland}

\author{Robert Fickler}
\email{robert.fickler@tuni.fi}
\affiliation{Tampere University, Photonics Laboratory, Physics Unit, Tampere, FI-33720, Finland}

\begin{abstract} 
Two-photon interference is a fundamental quantum optics effect with numerous applications in quantum information science.
Here, we study two-photon interference in multiple transverse-spatial modes along a single beam-path. 
Besides implementing the analogue of the Hong-Ou-Mandel interference using a two-dimensional spatial-mode splitter, we extend the scheme to observe coalescence and anti-coalescence in different three and four-dimensional spatial-mode multiports. 
The operation within spatial modes, along a single beam-path, lifts the requirement for interferometric stability and opens up new pathways of implementing linear optical networks for complex quantum information tasks.
\end{abstract}

\maketitle

Two-photon interference at a beamsplitter, i.e. Hong-Ou-Mandel (HOM) interference \cite{HOM_orig}, is one of the most important effects in photonic quantum information science \cite{bouchard2020twophoton}. 
Its applications range from quantum computing \cite{kok2007linear}, to cryptography \cite{xu2020secure} and from repeaters \cite{muralidharan2016optimal} to sensing \cite{giovannetti2011advances}, as well as quantum foundations \cite{bouwmeester1997experimental}. 
Due to its importance, it has been studied with photons from different sources \cite{somaschi2016near,he2019coherently} and in different degrees of freedom (DOF) \cite{Branning1999, Mohanty2017, Kobayashi2016}.
Domains that can encode high-dimensional quantum states, such as spatial, spectral, and temporal DOF, have attracted a lot of attention as they can be used to implement schemes with multiple input and output ports, i.e. multiports. 
Such linear optical networks are of importance for performing increasingly complex tasks in photonic quantum computing that require multi-photon interference \cite{carolan2015universal,brandt2020high,reimer2016generation,imany2019high,leedumrongwatthanakun2020programmable}. \par

Transverse-spatial modes, i.e. propagation invariant photonic structures that discretize the transverse-spatial domain, comprise a popular Hilbert space for encoding high-dimensional quantum states \cite{erhard2018twisted}. 
One prominent family of spatial modes is the Laguerre-Gaussian (LG) family that is defined by two quantum numbers, $\ell$ and $p$, describing the photons azimuthal and radial structures, respectively. 
The azimuthal DOF has gained significant popularity as it is connected to the orbital angular momentum (OAM) of photons \cite{Allen1992}.
Benefits of encoding quantum states in photonic spatial structures include mature technologies for generating and detecting high-dimensional states, as well as intrinsic phase stability of complex superposition states, ensured by single beam-path operation.

\begin{figure*}[htb] 
    \centering
    \includegraphics[width = \textwidth]{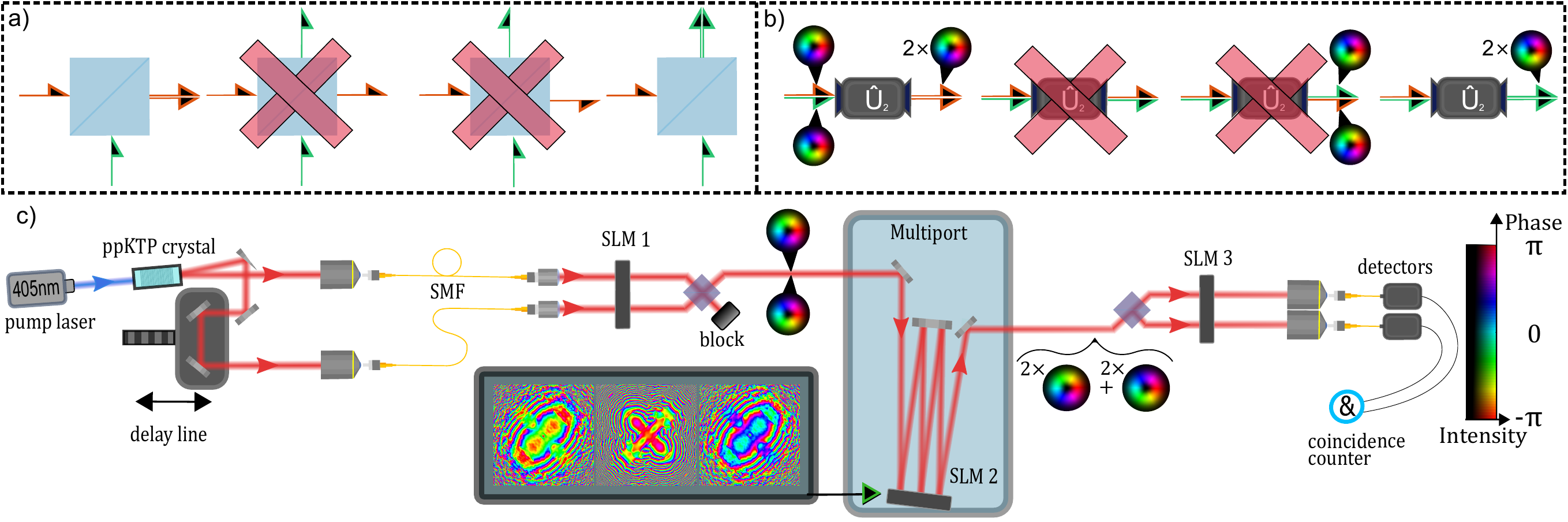}
    \caption{Conceptual sketch of two-photon interference and the used experimental setup.  a) Conventional HOM-interference in a regular beamsplitter between two paths. b) Its spatial mode analogue 
    implemented with a modesplitter.  
    c) The setup used to demonstrate two-photon interferences in different modesplitters. 
    Photon pairs are produced in a ppKTP (periodically poled potassium titanyl phosphate) crystal, adjusted in their  temporal overlap using a delay line , and coupled into single mode fibers (SMF).
    Three spatial light modulators (SLM) are used for spatial mode generation \cite{bolduc2013exact} (SLM1), unitary transformation \cite{brandt2020high} (SLM2), and measurement \cite{mair2001entanglement,krenn2017orbital} (SLM3).
    Single photon detectors and a coincidence counter are used to detect the photon pairs.
    For more details, see the main text and Supplementary.
    The two-dimensional colormap shown in c) is used in all of the figures.}
    \label{fig:Fig1}
\end{figure*}

In this article, we demonstrate two-photon interferences in multiple transverse-spatial modes.% along a single optical path. 
We use the technique of multi-plane light conversion (MPLC) \cite{labroille2014efficient} to implement various spatial-mode unitaries, leading to different interference effects between two structured photons.
We first observe bunching of photon pairs into the same spatial mode by implementing the direct spatial-mode analogue of HOM-interference using a two-dimensional ``modesplitter". 
Utilising the same setup and benefiting from the advantages spatial modes offer, we study various two-photon interferences in high-dimensional state spaces along with complex superpositions states.
The flexibility of our high-dimensional spatial-mode multiport further allows observing both coalescence and anti-coalescence of photon pairs with three and four input and output modes. 
Our demonstration opens up paths to realize novel implementations and complement existing high-dimensional linear optical networks for quantum information science.

In the conventional HOM-interference, photon bunching is obtained when two photons that are indistinguishable, i.e. perfectly overlapping in polarization, spatial structure, and time, are sent into a balanced beamsplitter from separate inputs.
While classically four different output situations are possible, only the two possibilities in which both photons exit through the same output-port will remain after the interference (see Fig. \ref{fig:Fig1}a), which can be attributed to the bosonic nature of photons.
A common way of assessing the quality of the interference is evaluating the change in coincident detections of the two exiting photons while scanning the temporal delay between them.
This quality can be quantified with a visibility $V = \frac{R_{cl}-R_{qu}}{R_{cl}} \in [0,1]$ \cite{Weihs1996} between the classically expected rate $R_{cl}$ and the rate $R_{qu}$ observed due to quantum interference.

In our experiment, we replace the beamsplitter acting on the paths with a modesplitter acting on the transverse-spatial modes of the photons.
Hence, the photons bunch into the transverse-spatial modes, which is in contrast to previous quantum interference measurements, where spatial-mode overlap served as a \textit{condition} for observing two-photon bunching into paths \cite{nagali2009optimal,Karimi2014,Zhang2016,malik2016multi,zhang2016engineering}. 
We note that in one recent experiment single-path two-photon interference between two spatially structured photons was observed, however, using polarization structures which limit the dimensionality to two \cite{d2019tunable}. 

To study HOM-interference between two spatial modes, we prepare the photons in orthogonal OAM-modes, i.e. one photon with $\ell = +1$ and the second having $\ell = -1$.
Note that although our scheme would be able to transform any combination of spatial modes \cite{brandt2020high}, we are only involving the OAM degree of freedom.
Using the Fock-state notation $\ket{n}_{\ell}$, where $n$ is the photon number and the subscript labels the OAM value, we can write the input state as $\ket{1}_{-1}\ket{1}_{+1} =\ket{1,1}$.
Analogous to the classic HOM-interference, a balanced modesplitter unitary $\hat{U}_2$ transforms the two input modes into two equally weighted superpositions which leads, via interference, to the state
\begin{equation}
\label{eq:2D_final}
    \ket{\Psi_{2D}} = \frac{1}{\sqrt{2}}\left(\ket{0,2} -\ket{2,0}\right),
\end{equation} 
if the two photons are perfectly indistinguishable in polarization, time, spectrum, as well as path (see Fig. \ref{fig:Fig1}b) and Supplementary for more details).
This state is distinct from earlier spatial-mode interference experiments %\cite{nagali2009optimal,Karimi2014,Zhang2016,malik2016multi,zhang2016engineering},
as it is a so-called NOON-state 
%in a single optical path
, which is a desirable state for quantum metrology tasks due to its increased phase sensitivity \cite{giovannetti2004quantum}. 
The single beam-path operation is similar to quantum interference with polarization \cite{Branning1999,slussarenko2017unconditional}, however, with the much larger state-space that spatial modes offer.
Thus, enabling more complex unitary transformations and a pathway to building single-path linear optical networks.

\begin{figure}[htb] 
    \centering
    \includegraphics[width = 0.45\textwidth]{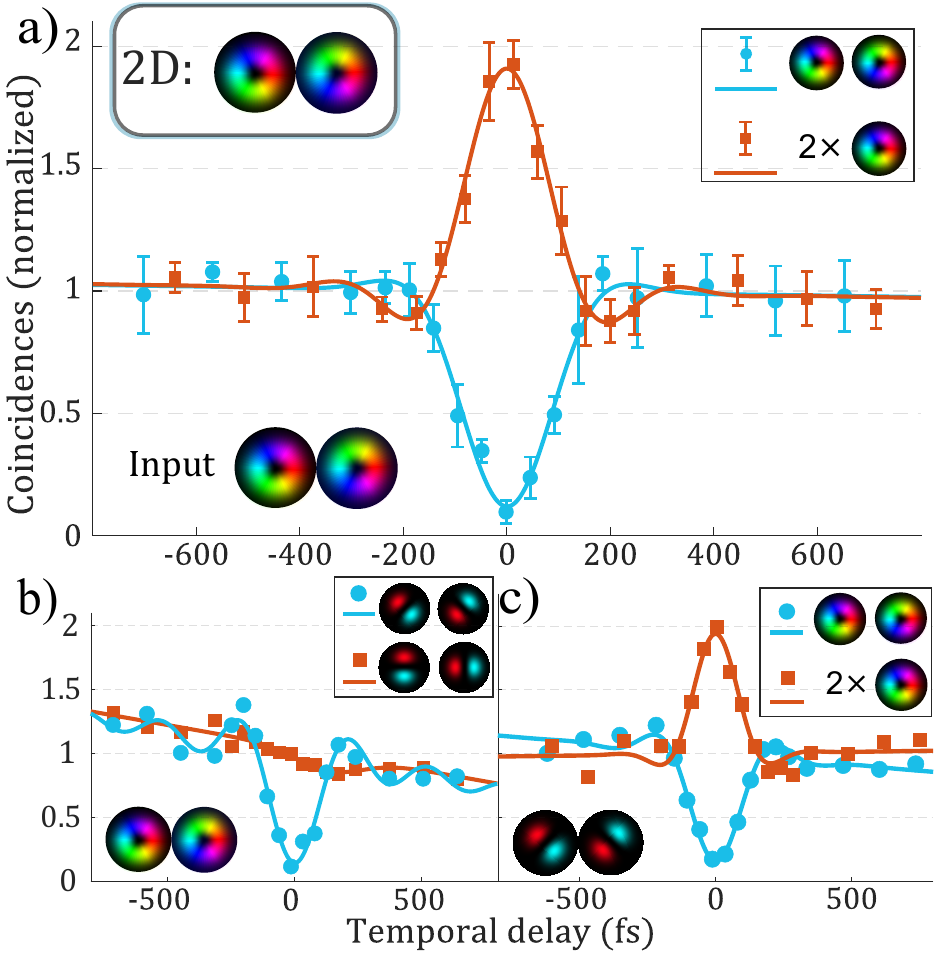}
    \caption{Measurement of two-photon bunching in a two-dimensional modesplitter.
    a) HOM-like interference visualized in Fig. \ref{fig:Fig1}b).
    Insets show the input modes of both photons (lower left) and legends depict the mode pair they were projected on (upper right). 
    b) Interference data with the same input two-photon state projected onto two different MUBs.
    c) Flipped scenario of b), i.e. the input states are in a different MUB and projected onto the OAM basis.
    The modesplitter unitary $\hat{U}_2$ was the same in all three scenarios. 
    The error-bars are standard errors calculated from multiple consecutive measurements and the curves are fits.
    In b) and c) the error-bars were omitted for clarity and can be found in the Supplementary.
    %\sugg{The change in the overall coincidence rates, especially visible in b), is a result of decoupling over time.}
    }
    \label{fig:2DResults}
\end{figure}

In the experiment (see sketch of the setup in Fig. \ref{fig:Fig1}c) and Supplementary), we generate photon pairs using spontaneous parametric down-conversion, ensure their temporal overlap through a delay line in the path of one photon, and couple both photons into single mode fibers (SMF), thereby spatially filtering them to a Gaussian mode. 
Using a fiber beamsplitter, we obtain an initial HOM-visibility of $97.7\%~\pm~0.2\%$ (see the Supplementary).
Note that for all visibility values given throughout the article, the errors denote a standard error calculated from a fit to the data.
%\sugg{For more information on the specific function we fit to the data, see the Supplementary.}
%\sugg{A few examples of these fits are shown in Fig. \ref{fig:2DResults}.}
We then imprint the desired spatial modes onto the photons using a spatial light modulator (SLM) and an amplitude and phase modulation technique to guarantee the best mode quality at the cost of loss (around $94-99\%$ loss per photon)~\cite{bolduc2013exact}. 
Lossless schemes exist \cite{hiekkamaki2019near} but we abstain from using them due to their complexity.
After imprinting the modes, the two photons are overlapped probabilistically with a balanced beamsplitter.
The input spatial modes are chosen to be orthogonal to avoid any interference in the combining beamsplitter.

The photons are then input into the spatial-mode multiport, i.e. an MPLC setup, that we use to implement any unitary operation on the bi-photon state. 
The three phase modulations in our MPLC setup, that define each unitary, are generated using free-space wavefront matching (WFM), which is described in more detail in the Supplementary and earlier works \cite{fontaine2019laguerre,brandt2020high}. 
When generating these transformations, the pixelization and limited number of phase values (8 bit) of our SLMs are taken into account.
Our 2D-modesplitter achieved a simulated efficiency above 99\% for a bandwidth of roughly 5 nm around the target wavelength.
Importantly, out of the 1\% loss, only a small fraction (1\%) remains in our operating state space, i.e. contributes to errors.
Note that this particular transformation for OAM-modes $\ell=\pm1$ resembles two rotated cylindrical lenses \cite{beijersbergen1993astigmatic}. 
The simulated evolution of the photon's structure in the 2D-transformation, with all the other utilized phase modulations, are displayed in the Supplementary.

After the two-photon state has been transformed through the modesplitter, the photons are probabilistically split into separate paths using another beamsplitter.
Their individual spatial modes are then measured using another phase modulation and SMF coupling followed by detection \cite{mair2001entanglement,krenn2017orbital}. 
The signals of the detectors are fed into a time tagging unit, which registers coincident detections through temporal correlations.
If the two-photon interference in spatial modes was successful, no coincidence counts will be detected when the photons are projected on orthogonal modes, i.e. using the projection operator $\hat{P}_{+1-1} = \ket{1}_{-1}\ket{1}_{+1}\bra{1}_{-1}\bra{1}_{+1}=\ket{1,1}\bra{1,1}$. 
Simultaneously scanning the temporal delay between the photons results in a dip in coincidence detections, identical to a classic HOM-dip.

For the 2D-modesplitter, we obtain a HOM-interference dip with a visibility $88.0\%~\pm~3.8\%$, that is well above the classical limit of $50\%$ \cite{rarity2005non}. 
When the two photons are projected onto the same state, i.e. $\hat{P}_{+1+1}=\ket{0,2}\bra{0,2}$, we observe an increase in coincidences due to bunching, i.e. a HOM-bump, with a visibility $90.9\%\pm4.5\%$.
This change in the projection only requires changing one hologram on SLM3.
Both results are shown in Fig. \ref{fig:2DResults}a).

We then take advantage of a particular benefit of spatial modes and study the interference when generating and detecting superposition states, a task that is usually difficult to implement in other degrees of freedom. 
At first, we keep the photons in the same input state, but project them onto orthogonal states of the two other mutually unbiased bases (MUB), which we define as $\ket{\Psi_{D/A}} = \frac{1}{\sqrt{2}}\left(\ket{1,0}\pm i\ket{0,1}\right)$ and $\ket{\Psi_{H/V}} = \frac{1}{\sqrt{2}}\left(\ket{1,0}\pm\ket{0,1}\right)$. 
When projecting the photons onto the orthogonal states of the (D/A)-MUB, we again find an interference-dip with a  visibility of $85.6\%~\pm~6.2\%$. 
However, when projecting both photons onto the (H/V)-MUB, no bunching is observed as both photons are transformed through the modesplitter into the eigenstates of this basis (see Fig. \ref{fig:2DResults}b)). 

To show that it is irrelevant whether we prepare or project onto superpositions, we then generate photons in the superposition states $\ket{\Psi_A}$ and $\ket{\Psi_{D}}$ but perform the projection measurements $\hat{P}_{+1-1}$ and $\hat{P}_{+1+1}$.
The results are similar to the ones using no superposition states, with visibilities $84.0\%~\pm~4.1\%$ and $93.8\%~\pm~9.5\%$, for the dip and the bump, respectively (see Fig. \ref{fig:2DResults}c)).
Note that, although the interference remains the same, the relative phase changes in the output state, allowing tuning of the obtained state.

To verify generation of the entangled state described in \eqref{eq:2D_final}, we perform an entanglement witness test on the two post-selected photons, which verifies non-separability if the sum of the visibilities measured in at least two MUBs is larger than 1~\cite{fickler2014quantum, Yu2005, guhne2009entanglement}.
From our measurements in all three MUBs, we obtain a witness value of $w = 2.2\pm0.1$, which is more than 11 standard deviations above the classical limit (see Supplementary for details).

\begin{figure}[htb]
    \centering
    \includegraphics[width=0.9\linewidth]{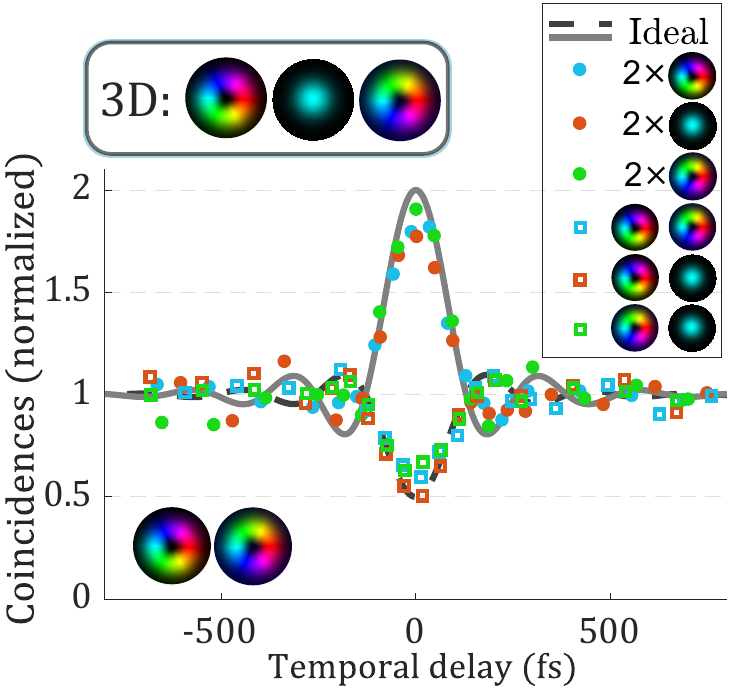}
    \caption{Two-photon interference in a three-dimensional modesplitter.
    The two-photon input state, shown in the inset on the lower left, was sent into a balanced unitary $\hat{U}_3$.
    The resulting interference was measured when projecting the photon pair onto every combination of our initial basis states.
    The ideal curves were calculated from the theoretically expected visibilities and the two-photon properties measured at the source.
    The omitted error-bars and more experimental details can be found in the Supplementary.
    }
    \label{fig:3D_resultsA}
\end{figure}

We then scale our state-space to larger dimensions, i.e. study two-photon interferences in  high-dimensional modesplitters.
At first, we choose a balanced three-dimensional modesplitter $\hat{U}_3$ operating on the state space spanned by LG-modes with $\ell=-1,0,+1$.
With this unitary, that splits the photons into an even superposition of all three spatial modes, different interference effects can be observed.
For example, if we again send in the same input state, now written as $\ket{\Psi}=\ket{1}_{-1}\ket{0}_{0}\ket{1}_{+1} = \ket{1,0,1}$, we won't observe a perfect bunching.
By measuring correlations using any of the projectors $\hat{P}_{+1-1}$, $\hat{P}_{+10}$, or $\hat{P}_{0-1}$, a HOM-dip with a maximum visibility of $0.5$ can be observed (see Supplementary).
Experimentally we measure visibilities of $39.7\%~\pm~2.9\%$, $49.3\%~\pm~3.1\%$, and $38.3\%\pm2.8\%$.
Similarly to the two-dimensional case, projecting both photons onto the same mode with $\hat{P}_{+1+1}$, $\hat{P}_{00}$, or $\hat{P}_{-1-1}$, results in a two-fold increase in coincidences.
The corresponding measurements lead to visibilities $84.3\%~\pm~5.6\%$, $77.2\%~\pm~4.8\%$, and $94.7\%~\pm~7.2\%$, respectively (see Fig. \ref{fig:3D_resultsA}).
%\sugg{For the high-dimensional interference curves, we display ideal curves instead of fits, to keep the figure simple and compare the data to theoretically expected results.}
The imperfect visibilities are likely due to small misalignments and imperfections in the unitary implemented with only three phase-modulations for  these increasingly complex transformation.
For completeness, we also confirmed that this three dimensional unitary works with superposition states, which is shown in the Supplementary.

\begin{figure}[t!] 
    \centering
    \hspace{-1cm}
    \includegraphics[width=0.8\linewidth]{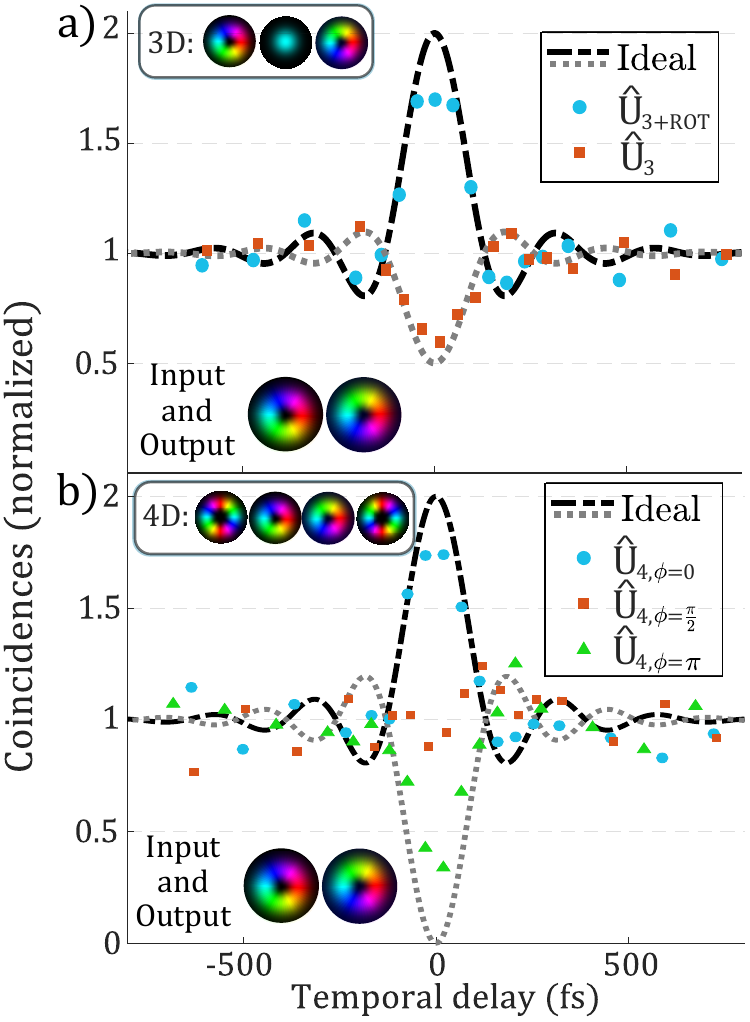}
    \caption{Two-photon anti-coalescence in a high-dimensional unitaries. 
    The results in a) show how imperfect coalescence in a three-dimensional state space can be changed to anti-coalescence by changing to an unbalanced modesplitter unitary. 
    Further, in b), three different balanced four-dimensional modesplitters are used to tune the observed coalescence with the same input and output states.
    Definitions for the unitaries and error-bars can be found in the Supplementary.
    }
    \label{fig:AntiCoal}
\end{figure}

Naturally, scaling these effects into higher-dimensional state spaces, i.e. realizing linear optical networks, is important for quantum information applications \cite{carolan2015universal,brandt2020high,reimer2016generation,imany2019high,leedumrongwatthanakun2020programmable}.
However, this scaling also provides some fundamentally interesting effects that can't be observed in a classical HOM setting, i.e. a two-dimensional system.
One such effect is anti-coalescence, where two-photon interference causes an increase in coincidences when projecting the bi-photon state onto two orthogonal spatial modes, while still having a separable state as an input.
In three-dimensions, this phenomena can be observed if either the input states or the unitary is prepared in an unbalanced superposition basis. 
An example of the latter has been demonstrated using paths and an imperfect tritter \cite{mattle1995nonclassical}.
We demonstrate anti-coalescence by exploiting the flexibility of our multiport, and compare the interference obtained with the balanced modesplitter $\hat{U}_3$ to an unbalanced unitary $\hat{U}_{Rot+3}$ while keeping the same two-photon state $\ket{1,0,1}$ as input and output.
As already shown in Fig. \ref{fig:3D_resultsA}, for the balanced modesplitter $\hat{U}_3$ we observe coalescence.
However, when using the unbalanced modesplitter $\hat{U}_{Rot+3}$ we find an increase in coincidences with a visibility of $77.4\%~\pm~6.4\%$ caused by anti-coalescence (see Fig. \ref{fig:AntiCoal}a).
Because of the imperfections outlined earlier, our measured visibilities are slightly lower than the theoretically expected 100\%.  
Due to the bosonic nature of photons, bunching is still the driving force of these interferences, which manifests as HOM-dips when projecting on the two other orthogonal mode pairs (see Supplementary).
We further verified the same anti-coalescence using the balanced unitary $\hat{U}_3$ and unbalanced superposition states, which can be found in the Supplementary.

When going beyond three-dimensions tuning the observed interference becomes easier since the modesplitter unitary can be kept balanced while changing it's internal phases \cite{mattle1995nonclassical}. 
We demonstrate this tunability using a balanced four-dimensional modesplitter
%\sugg{\begin{equation}
%    \hat{U}_4 = \frac{1}{2}
%    \begin{bmatrix}
%    1 & 1 & 1 & 1 \\
%    1 & e^{i\varphi} & -1 & -e^{i\varphi}  \\
%    1 & -1 & 1 & -1 \\
%    1 & -e^{i\varphi} & -1 & e^{i\varphi}
%    \end{bmatrix},
%    \label{eq:U4}
%\end{equation}}
for which we adjust the internal phase values $\varphi$ to $0$, $\frac{\pi}{2}$, or, $\pi$, corresponding to an anti-coalescence, no interference, and coalescence, respectively.
The chosen four-dimensional state space is spanned by the OAM-modes with $\ell = \pm2,\pm1$.
%\sugg{Although any two-mode combination of the mode set could have been used, we kept the same input state $\ket{\Psi} = \ket{0}_{-2}\ket{1}_{-1}\ket{1}_{+1}\ket{0}_{+2} = \ket{0,1,1,0}$ and projected on the same output state, i.e. $\hat{P}_{-1+1}$, as before to show that the change in interference only stems from the different unitary operation.}
In our measurements (shown in Fig. \ref{fig:AntiCoal}b)) we obtained a visibility of $75.0\%~\pm~6.1~\%$ for $\varphi=0$, no significant interference for $\varphi=\frac{\pi}{2}$ and a visibility of $63.2\%~\pm~6.4\%$ for $\varphi=\pi$.
While not being perfect, the obtained visibilities are above the classical limit, with at least $95\%$ confidence.
We again attribute the discrepancy between theory and experiment to the limitations of our small MPLC system performing more complex transformations.
Already in simulations, the limited number of phase screens leads to an increase in mode-independent loss, around $27\%-37\%$, and a slightly unbalanced modesplitter.
Although MPLC setups have been implemented using a larger number of phase modulations on a single SLM \cite{fontaine2019laguerre,brandt2020high}, we refrained from doing so here due to the additional losses induced by every SLM reflection.

The multi-photon interference effects shown here, demonstrate that a re-configurable spatial-mode multiport, implemented through MPLC, can be used in the quantum domain, opening up multiple new research avenues and quantum technological applications harnessing the benefits of spatial modes. 
The current limitation of our experimental scheme is the lossy method of generating the spatial modes \cite{bolduc2013exact} and the limited efficiency of our SLMs (75\% efficiency per reflection), that limits the number of phase modulation planes. 
However, these limitations are only of technical nature and can be tackled in the future with more expensive devices, e.g. high-quality deformable mirrors, and novel methods, e.g. lossless generation and detection of structured photons \cite{hiekkamaki2019near}. 
Because our scheme is intrinsically stable and can be fully automized \cite{brandt2020high}, scaling to large mode numbers, i.e. the realization of large linear optical networks along a single path, seems feasible.
%\sugg{Additionally, other input states, such as entangled states or multi-partite states, could be used to investigate more complex multi-photon interferences.}
The well-controlled two-photon interference can further be applied in generating custom-tailored NOON states of spatial modes, studying complex quantum walks within the spatial-mode set \cite{grafe2016integrated}, simplifying fundamental research endeavours such a high-dimensional multi-partite entanglement \cite{erhard2018experimental}, or applying spatial modes in complex quantum information tasks like photonic quantum processors \cite{peruzzo2014variational,carolan2015universal}, high-dimensional quantum teleportation \cite{luo2019quantum}, or Boson sampling \cite{wang2017high}.  

The authors thank Marcus Huber, Mario Krenn, Shashi Prabhakar, and  Lea Kopf for fruitful discussions.
MH and RF acknowledge the support of the Academy of Finland through the Competitive Funding to Strengthen University Research Profiles (decision 301820) and the Photonics Research and Innovation Flagship (PREIN - decision 320165). MH also acknowledges support from the Magnus Ehrnrooth foundation through its graduate student scholarship.

\bibliography{Biblio.bib}

\end{document}

\title{Supplementary to: \\ High-dimensional two-photon interference effects in spatial modes}

\author{Markus Hiekkam\"aki}
\email{markus.hiekkamaki@tuni.fi}
\affiliation{Tampere University, Photonics Laboratory, Physics Unit, Tampere, FI-33720, Finland}

\author{Robert Fickler}
\email{robert.fickler@tuni.fi}
\affiliation{Tampere University, Photonics Laboratory, Physics Unit, Tampere, FI-33720, Finland}

%\flushbottom
\maketitle
\renewcommand\thefigure{S\arabic{figure}}    
\setcounter{figure}{0} 
\renewcommand\theequation{s\arabic{equation}}    
\setcounter{equation}{0}

%\section*{Supplementary} 
\label{sec:suppl}

\section*{Two-photon interference in the spatial mode domain}
To compute the different output states from two-photon interference, we need to first define our states and creation operators. 
In the main article we denote LG-mode Fock-states as $\ket{n}_{\ell}$, where $n$ is the photon number and the subscript labels the OAM value.
We can then define the creation operator $\hat{a}_{\ell}^{\dagger}$ for a mode carrying $\ell$ quanta of OAM (in path $a$), such that we can write our two-dimensional input state as $\hat{a}_{-1}^{\dagger}\hat{a}_{+1}^{\dagger}\ket{0}_{-1}\ket{0}_{+1}=\ket{1}_{-1}\ket{1}_{+1} = \ket{1,1}$.
In our two-dimensional spatial mode HOM-interference, the unitary performed by the modesplitter transforms the two input modes into two equally weighted superpositions, i.e. it is of the form
\begin{equation}
\label{eq:2DUnitary}
    \hat{U}_2 = \frac{1}{\sqrt{2}}
    \begin{bmatrix}
    1 & -1 \\
    1 & 1
    \end{bmatrix}.
\end{equation} 
With this unitary the input OAM-modes are turned into first order Hermite-Gaussian modes, where a scheme using cylindrical lenses is also known \cite{beijersbergen1993astigmatic}.
When we input the photons in the state $\ket{1,1}$ into the modesplitter, we obtain the state
\begin{equation}
\label{eq:2D_HOM}
\begin{aligned}
    \ket{\Psi_{2D}} = & \hat{U}_{2}\ket{1,1}\\
    = & \hat{U}_{2}\hat{a}_{-1}^{\dagger}\hat{a}_{+1}^{\dagger}\ket{0,0} \\
    = & \frac{1}{2}\left(\hat{a}_{-1}^{\dagger}\hat{a}_{-1}^{\dagger} +  \hat{a}_{-1}^{\dagger}\hat{a}_{+1}^{\dagger}-\hat{a}_{+1}^{\dagger}\hat{a}_{-1}^{\dagger} -\hat{a}_{+1}^{\dagger}\hat{a}_{+1}^{\dagger}\right)\ket{0,0}.
\end{aligned}
\end{equation}
If the two photons are perfectly indistinguishable in polarization, time, as well as path, the two middle terms destructively interfere such that we obtain the state
\begin{equation}
\label{eq:2D_final}
    \ket{\Psi_{2D}} = \frac{1}{\sqrt{2}}\left(\ket{0,2} -\ket{2,0}\right).
\end{equation}
We see, that both photons leave the modesplitter either with $\ell = +1$ or with $\ell = -1$, as expected. 
Projecting this bi-photon state onto a state where the two photons have orthogonal spatial modes, i.e. a measurement corresponding to the projection operator $\hat{P}_{+1-1} = \ket{1}_{-1}\ket{1}_{+1}\bra{1}_{-1}\bra{1}_{+1}=\ket{1,1}\bra{1,1}$, one should not be able to detect any coincidence counts.

%\sugg{To compare this N00N-state to two-photon states produced in previous quantum interference experiments involving spatial modes \cite{nagali2009optimal,Karimi2014,Zhang2016,malik2016multi,zhang2016engineering}, the resulting quantum state in these aforementioned experiments is
%of the form $\frac{1}{2}\left(\hat{a}_{SM1}^{\dagger 2} - \hat{b}_{SM2}^{\dagger 2}  \right)\ket{0,0}$, where $a$ and $b$ are the two orthogonal paths, SM1 is some spatial mode, and SM2 is an orthogonal spatial mode.
%Here, each spatial mode only exists on the photons bunching into the corresponding path, i.e SM1 only exits on photons in path a and vice versa.
%In contrast, in the same operator notation, we generate a state of the form $\frac{1}{2}\left(\hat{a}_{SM1}^{\dagger 2} - \hat{a}_{SM2}^{\dagger 2}  \right)\ket{0,0}$, i.e. both of the output options exist in the same path $a$, instead of being split between the paths $a$ and $b$.}

\section*{Wavefront matching}
Wavefront matching is the computerized optimization algorithm that created each set of phase modulations that we use in our modesplitters.
In a simplified picture, this optimization takes a set of input modes, and the corresponding output modes we want to convert the input modes into.
The protocol then simulates propagation, of each input mode, through the MPLC system (where initially the phase elements are blank).
The same simulation is also done backwards through the system for the desired output modes.
After these simulations, the overlap between each backwards propagated output mode and the corresponding forwards propagated input mode is calculated at the position of one phase element.
These overlaps at the phase element are then used to create a structure for the phase element, in such a way that the differences between the input and output mode are minimized at this plane.
(For a more detailed description of this optimization algorithm, and optimization function, see the original free-space WFM article \cite{fontaine2019laguerre}, or one of our previous articles \cite{brandt2020high, hiekkamaki2019near}).
\begin{figure}[t!]
    \centering
    \includegraphics[width = \linewidth]{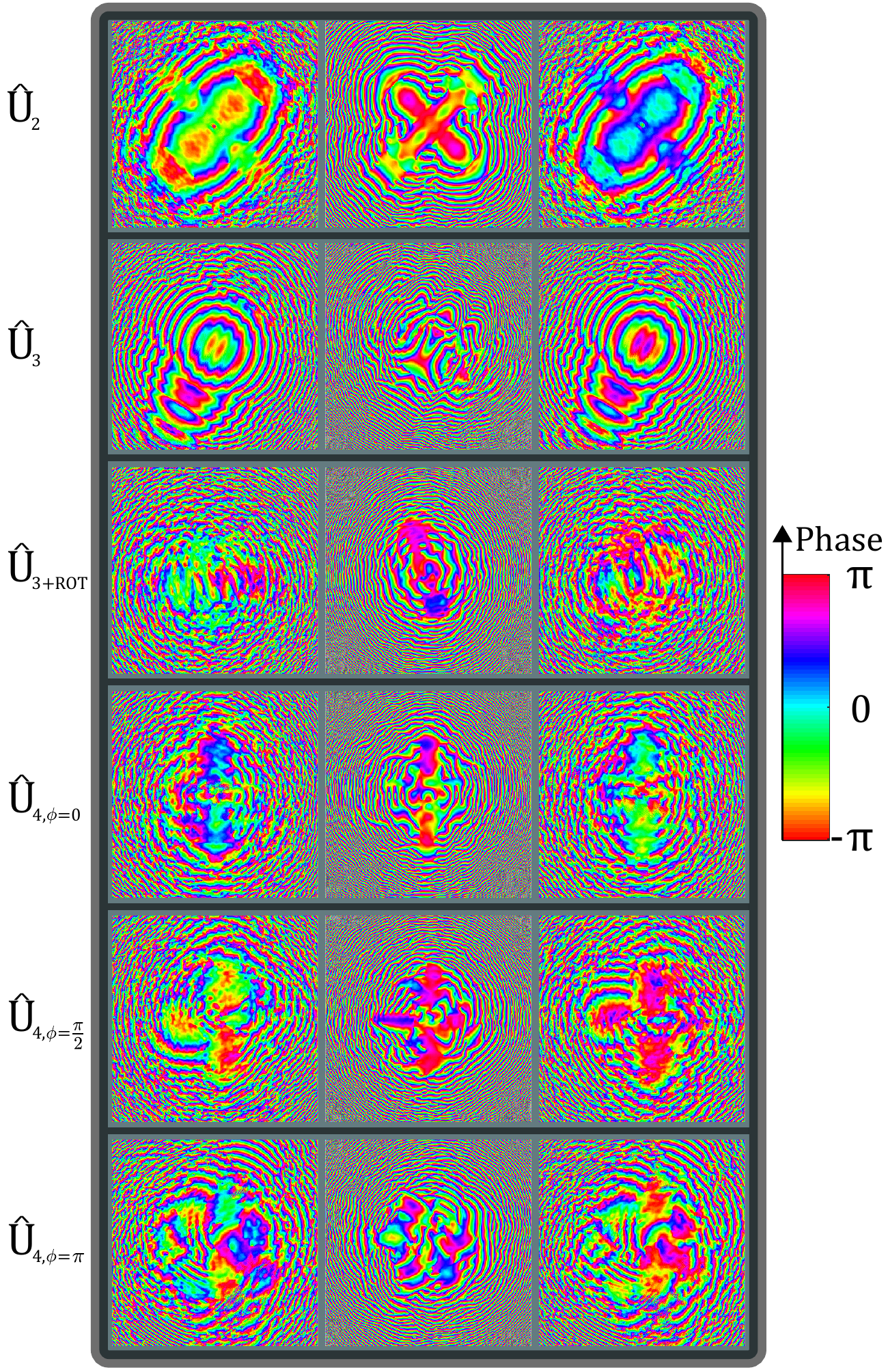}
    \caption{All of the phase-modulation screens from the transformations we used, along with the corresponding unitaries. The unitaries are displayed in the order they appear in the main article.}
    \label{fig:Trafos}
\end{figure}
After the structure of this phase screen has been optimized once, we update the propagation simulations and update the next phase element similarly.
This process is then repeated for each phase element in the system, sequentially.
The whole procedure can be repeated multiple times for all phase elements until a sufficient quality is achieved.
For our devices, we performed around 100 repetitions to obtain sufficiently good transformations.
We also included multiple wavelengths into the same optimization algorithm, in order to achieve a good quality of the transformations for our photon pairs that had a wide bandwidth (see following section).
An example of our wavefront matching code can be found online \cite{hiekkamaki_markus_2019_3570622}, which
only works for one mode at a time, but can easily be scaled to multi-mode operations.
All of the phase modulation screens we used in the main article are displayed in Fig. \ref{fig:Trafos}.
As an example, the transformation performed by our two-dimensional unitary is shown in Fig. \ref{fig:2DBlackbox}.
\begin{figure*}[t]
    \centering
    \includegraphics[width = \textwidth]{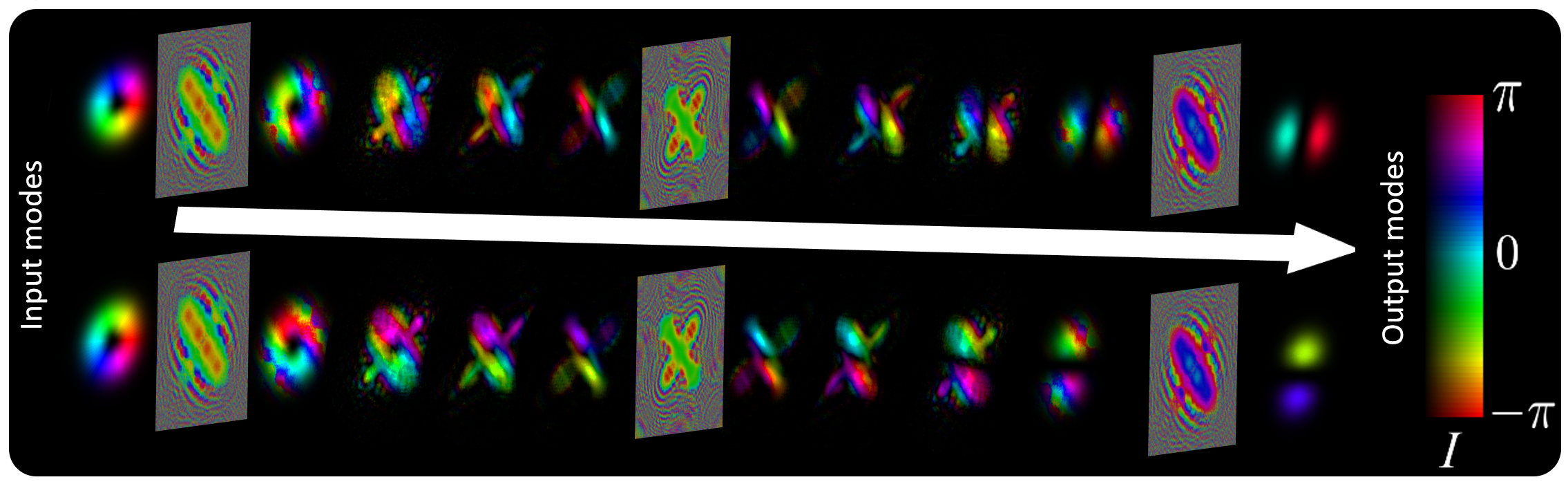}
    \caption{Transformation showing how both OAM-modes (from the two-dimensional state space) are transformed into an equal superposition of the same two modes.
    The displayed transformation corresponds to the unitary $\hat{U}_2$. 
    The output modes in this particular case correspond to the first order Hermite-Gaussian modes.
    The slight variations in the phase modulation screens, when comparing to the displayed transformation in Fig. \ref{fig:Trafos} (used in experiments), are due to minor differences in the parameters used for WFM.}
    \label{fig:2DBlackbox}
\end{figure*}

\section*{Details of the experimental setup}
\begin{figure*}[tbh]
    \centering
    \includegraphics[width=\linewidth]{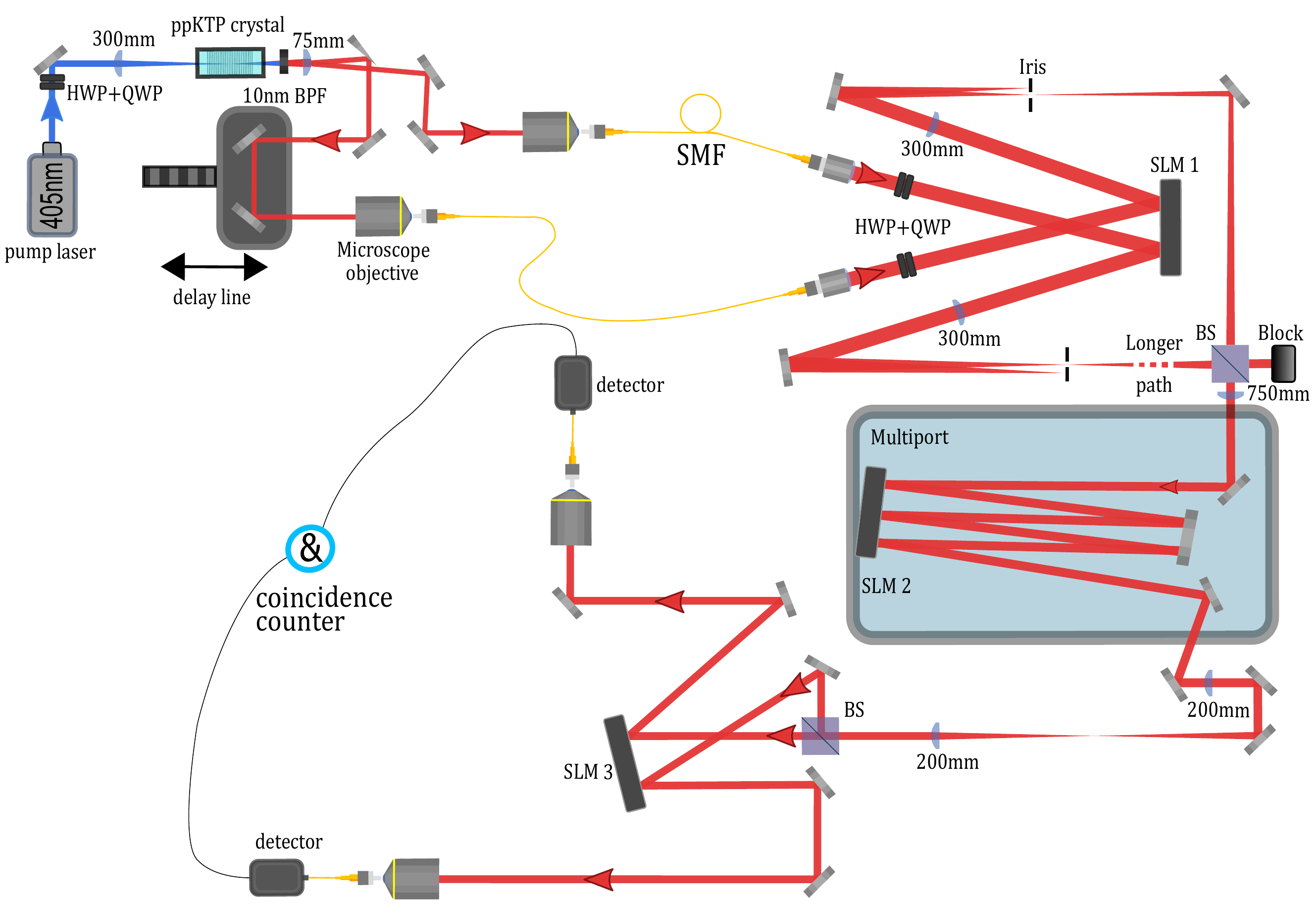}
    \caption{A detailed drawing of the experimental setup. 
    The two-photons are generated in the top left where a 405nm pump is focused into a ppKTP crystal (type-0, 12~mm). The polarization of the pump is optimized for the crystal with one half-wave plate (HWP) and one quarter-wave plate (QWP). The pump is then removed with a 10nm bandpass filter (BPF) and the two photons are separated from each other using a lens that splits them according to their momentum correlations.
    One of the two photons then passes through an adjustable delay line and both of the photons are coupled into single mode fibers (SMF). 
    After this the photons are sent to the first spatial light modulator (SLM1, Holoeye Pluto 2) where any spatial mode can be imprinted on each photon. 
    A 4f-system is then used to select only the first diffraction order from SLM1, since the method used for generation is holographic. 
    In the middle of this magnifying 4f-system (magnified such that we can use a larger area of SLM2), the two photons are overlapped using a beamsplitter and after reimaging, they are sent to the device peforming the unitary. 
    This device has been already described in the main article.
    After the unitary device, the two photons are split and reimaged onto the third SLM on which a holographic method is used to measure the spatial mode on each of them. 
    In this measurement, both photons are coupled into SMF's and detected using avalanche photodiode detectors.
    The correlations of these detections are time-tagged using a coincidence counter in order to distinguish which photons belong to which pair.}
    \label{fig:detail_setup}
\end{figure*}

\textbf{Photon pair source}: The photon pairs are created in a 12~mm type-0 ppKTP crystal that we pump with a narrow band 405~nm laser to achieve spontaneous parametric downconversion.
Our pump laser power varied between 40mW and 70mW due to a slow degradation over time.
Due to this degradation of our laser and it's output spatial mode, the source output varied between 24 to 40 million single photons with a coincidence to singles ratio of 6 to 15~\%.
The photons were split using their momentum anti-correlations, by inserting a 2f-system, that sends each half of the pair into opposite sides of the Fourier-plane, where a D-shaped mirror is put in half of the beam to reflect only one photon (see Fig. \ref{fig:detail_setup}).
The produced photon pairs are degenerate at 810~nm and we filter them with a 10nm bandpass filter, that has a square wavelength profile, before coupling them into single-mode fibers.
The source also includes an adjustable delay line for one of the photons, that we can use to control the temporal indistinguishability of the photon pairs.
All of these details can be seen from a sketch of the experimental setup shown in Fig. \ref{fig:detail_setup}.\\
From the bi-photon spectrum, we expect roughly a $\sinc$-like HOM-interference profile when the temporal overlap is the degree of freedom being scanned  \cite{Prabhakar2018}.
However, due to spectral filtering, when coupling the photons into single-mode fibers, and due to multiple blazed gratings along the whole beam-line, we expect a slight change in the spectral profile of the photons which we model as a Gaussian attenuation.
Thus, the curve we fit to all of our data was
\begin{equation}
\label{eq:general_Dip}
R_c = R_{cl} \left( 1\pm V \sinc \left(d_1\Delta t\right)\exp\left(-(d_2\Delta t)^2\right) \right) + S\Delta t,
\end{equation}
where $d_1$ and $d_2$ are constants that depend on the spectral profile of the photons, $S$ is a slope that takes into account any linear decoupling in the scanning processes, $R_{cl}$ is the classically expected coincidence rate, $\Delta t$ is the time delay between the photons and $V$ is the visibility. 

\textbf{Mode generation}: In our setup we generate the separable bi-photon spatial mode state using one amplitude and phase modulating hologram (on SLM1), for each photon.
These holograms are able to generate any complex spatial mode structure by carving out the required amplitude shape and imprinting the corresponding phase pattern \cite{bolduc2013exact}. 
However, as described in the main text, this modulation method induces a lot of loss if used to generate the best possible spatial modes. 
In addition, due to the carving process, the losses depend on the size of the mode structure, i.e. the smaller the mode structure the larger the losses.

\textbf{Mode manipulation}: The multi-plane light conversion (MPLC) was realized through three consecutive phase modulations with, in our case, a propagation of 0.8~m between each of the modulations. 
%The free-space propagation is required to enable a lossless amplitude alteration through phase modulations only. 
%In the Supplementary, we show the simulated beam evolution for each input beam in the case of the unitary \eqref{eq:2DUnitary} and the input modes with OAM $\ell=\pm1$. 
%The required phases are found using the technique of wavefront matching, which is explained in more detail in the Supplementary and also in earlier works\cite{fontaine2019laguerre,hiekkamaki2019near,brandt2020high}.\\

\textbf{Mode measurement}: Our spatial mode measurements start by probabilistically splitting the two photons with a balanced beamsplitter, which successfully splits the photons with a 50\% probability.
Then, the spatial modes of both photons are measured individually through phase-flattening holograms and single-mode fibers (SMF) \cite{krenn2017orbital}.
In this measurement scheme, the holograms flatten the phase-structure of a specific spatial mode.
As the SMF only efficiently couples photons that have a flat phase, the combination of a flattening hologram and an SMF acts as a spatial mode filter. 
The complete bi-photon state is then measured by performing the above described projective measurement simultaneously on each of the two photons, and detecting both photons with two avalanche photodiode detectors while correlating the measurements recorded within a coincidence window of 1 ns through a time tagging unit.

The detectors we used were Laser components Count-T avalanche photodiode detectors and the time tagging unit was an ID Quantique ID900.
All of the SLMs we used were Holoeye's Pluto 2 reflective spatial light modulators.
Note, that the beamsplitter used for combining (measuring) the two photons could be replaced with spatial mode sorters that multiplex (de-multiplex) these spatial modes into different paths \cite{fontaine2019laguerre}.
When using such a scheme in combination with photon number resolving detectors, optimal efficiencies could be achieved.

\section*{Photon pair source visibility}
\begin{figure}[t]
    \centering
    \includegraphics[width=\linewidth]{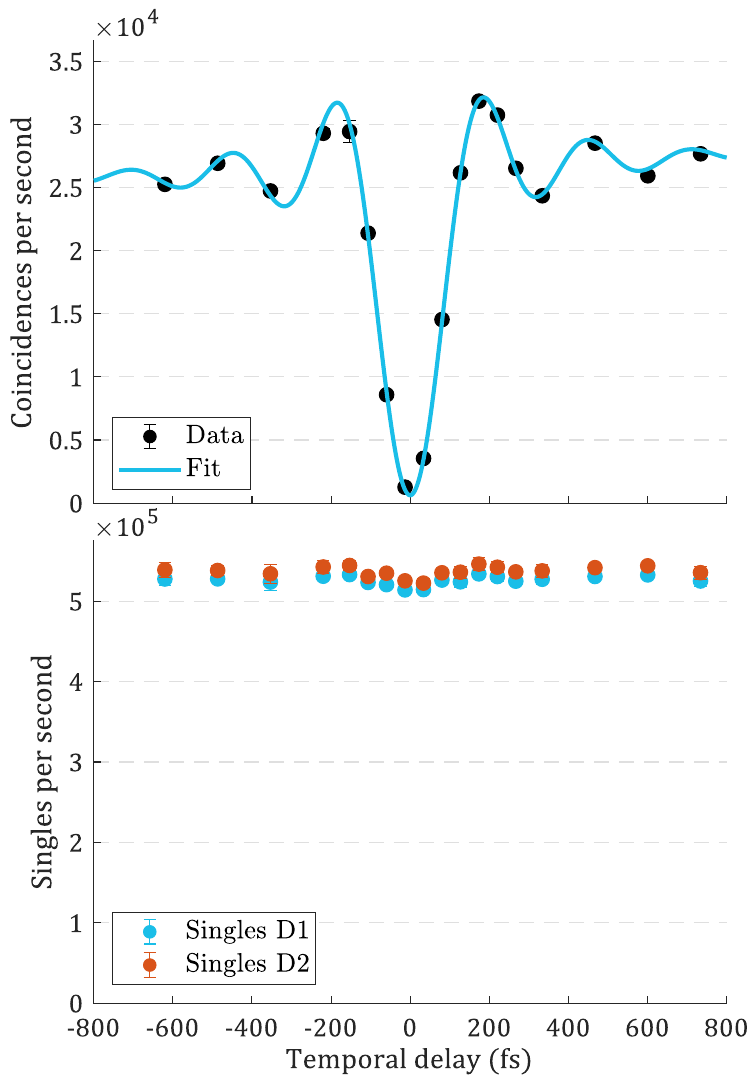}
    \caption{ HOM measurement characterizing our photon pair source.
    The measurement was done by sending the pairs through a 50/50 fiberbeamsplitter.
    The visibility of the recorded HOM-dip was $97.7 \%\pm 0.2 \%$.
    Accidental coincidences have been removed from the data.
    The dip was measured with lower power (approx. 1.5 mW) due to the finite dead time of our detectors.
    In the second plot, we show the corresponding single-photon rates.
    We see a decrease in single-photon detections at 0 delay due to the increase in $\ket{2}$ Fock-states, that our single-photon detectors register as single-photon detections.}
    \label{fig:Initial_dip}
\end{figure}

We measured the quality of our photon pair source in a regular HOM-interference experiment, where we combined our two photons in a fiber beamsplitter.
When scanning the temporal overlap between the photons, we achieved a HOM-dip with a visibility of $97.7 \%\pm0.2 \%$.
To achieve this visibility, we have removed the accidental coincident detections that we calculate with
\begin{equation}
\label{eq:acc}
    Acc = S_1 * S_2 * t_{window},
\end{equation}
where $S_1$ and $S_2$ are the measured single-photon rates and the coincidence window $t_{window}$ was 1ns.
The initial HOM-dip can be seen in Fig. \ref{fig:Initial_dip}.\par

 \par

With the initial HOM-dip, we see a decrease in photon rates in the corresponding single-photon data.
A decrease in single-photon events, which is 50\% of the decrease in coincidence events, is expected with our single-photon detectors that are not photon-number resolving.
This decrease stems from the increased number of $\ket{2}$ Fock-states in each arm when the photon pairs start bunching.
Since these $\ket{2}$-states are divided evenly between the two outputs one observes a reduction in the single-photon rates, in both arms, that is 50\% of the reduction in coincidence counts.
When fitting the same interference function into our single-photon data, we noticed a decrease in single-photon detections that was 56.8\% of the decrease in coincidence detections.
The remaining 6.8\% are assumed to be due to the dead time of our detectors.

\section*{Processing of the experimental data and displayed errors}
\begin{figure}[htb]
    \centering
    \includegraphics[width=\linewidth]{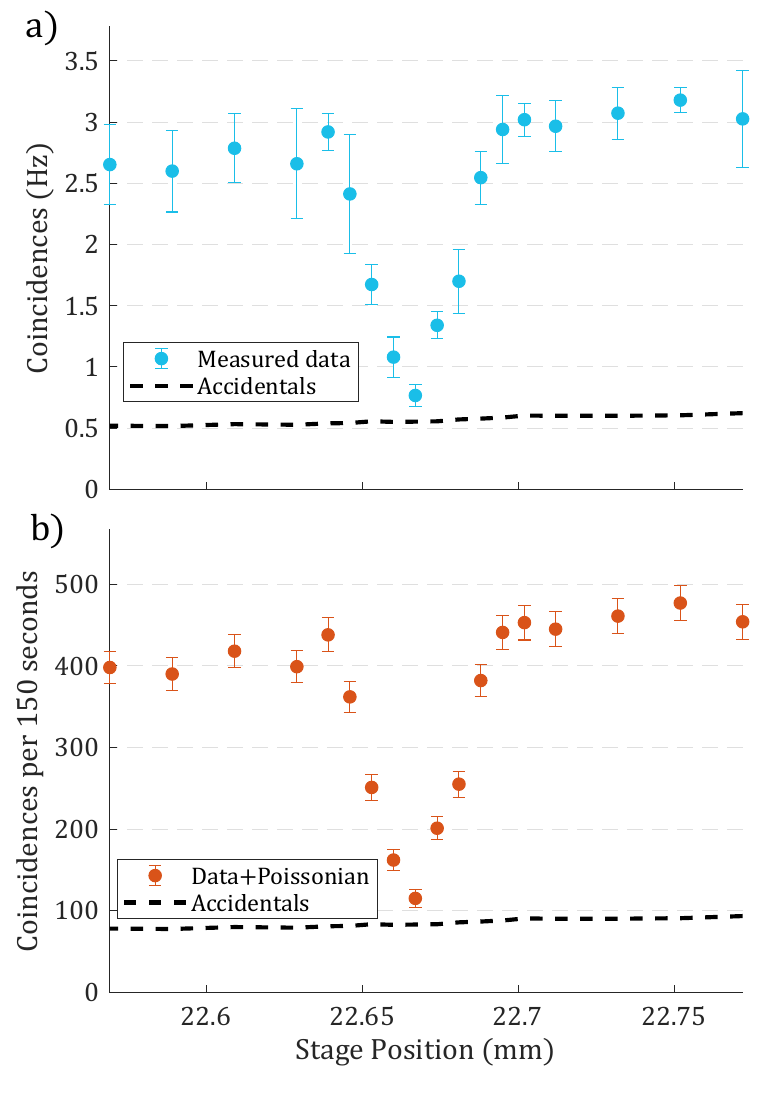}
    \caption{ Example of the raw data measured for HOM-interference between two spatial modes.
    The data in plot a) is the same as the one displayed in Fig. 2a) of our main article.
    The standard errors in a) were calculated from five repetitions of the measurement.
    For comparison, in b) the errorbars are estimations of the standard error, based on Poissonian statistics and the total number of coincidences we measured at each position of our delay line.
    The reason for why our measured errors are bigger than the Poissonian errors (which is the minimum achievable error with our probabilistic photon pair source and losses) is most likely due to instabilities in the position of the delay line stage and local thermal instabilities in the SMFs and the ppKTP crystal.}
    \label{fig:Raw_dip}
\end{figure}

As we referred to in the main article, each interference measurement was corrected for accidental detections using equation \eqref{eq:acc}.
We also corrected for some effects caused by decoupling or degradation of our source. 
These effects manifested as a decrease in single-photon rates over time.
One example can be seen in the raw data presented in Fig. \ref{fig:Raw_dip}.
The figure displays the raw data corresponding to the first HOM-dip we showed in the main article.
Since we scanned our delay stage backwards, the decoupling emerges as a linear increase in the single-photon events (from left to right, in the plot).
We corrected for this effect by taking the average single-photon rate over the whole scan and the percentile change in single-photon events at each position.
We used these percentage values to correct for the small linear decoupling in our measured data. \par
This small change in single-photon detections is reflected in the change in accidental detections shown in Fig. \ref{fig:Raw_dip}.
Fig. \ref{fig:Raw_dip} also compares the measured data to how well an estimate of Poissonian errors accounts for our experimental errors.
Furthermore, since the error-bars were omitted when plotting our results in three and four-dimensional unitaries, we have plotted all of these measurement results, with error-bars, in Fig. \ref{fig:HD_rawdata}.

\section*{Entanglement Witness}
The state \eqref{eq:2D_final}, which is generated in the two-dimensional modesplitter operation, can also be rewritten in the OAM basis, instead of the Fock-state basis. When doing so, we obtain
\begin{equation}
    \ket{\Psi_{2D,OAM}} = \frac{1}{\sqrt{2}}\left(\ket{-1,-1}-\ket{+1,+1}\right),
    \label{eq:2d_OAM_ENT}
\end{equation}
which corresponds to a qubit entangled state. 
Hence, we can probabilistically split the two photons into two separate paths, post-select on the events when this separation was done successfully, and use an entanglement witness to study the obtained entanglement.
In our entanglement witness measurements, we measured coincidences at zero delay for 12 different measurement combinations.
%The input states were OAM modes with $\ell = \pm 1$, for all measurements, and the unitary was the two-dimensional unitary $\hat{U}_2$ we introduced in the main article.

We measured all four permutations of the $\ell = \pm 1$ OAM-modes on the bi-photon state as well as all four permutations of basis states in each of the two other MUBs that were introduced in the main article.
From these 12 measurements we calculated the entanglement witness value $w$, which is \cite{fickler2014quantum, Yu2005, guhne2009entanglement}
\begin{equation}
    \label{eq:witness}
    w = |\left\langle\hat{\sigma_x} \otimes \hat{\sigma_x}\right\rangle| + |\left\langle\hat{\sigma_y} \otimes \hat{\sigma_y}\right\rangle| + |\left\langle\hat{\sigma_z} \otimes \hat{\sigma_z}\right\rangle|.
\end{equation}
This value simplifies to 3 visibilities in the three available MUBs, and the witness is a sum of these three visibilities.
If the sum exceeds the value of 1, the state is non-separable. In our experiment, we obtain a witness value of we obtained a value of $w = 2.2\pm0.1$, which is more than 11 standard deviations above the classical limit. 

\section*{Three-dimensional states and unitaries}

For the initial input state $\ket{1,0,1}$ into our balanced three-dimensional unitary
\begin{equation}
\label{eq:U3}
    \hat{U}_3 = \frac{1}{\sqrt{3}}
    \begin{bmatrix}
    1 & e^{i\frac{4\pi}{3}} & e^{i\frac{2\pi}{3}} \\
    1 & e^{i\frac{2\pi}{3}} & e^{i\frac{4\pi}{3}} \\
    1 & 1 & 1
    \end{bmatrix},
\end{equation}
where the first, second, and third columns correspond to $\ell = -1,0,$ and $+1$, respectively , the theoretical output state was
\begin{equation}
\begin{aligned}
    \ket{\Psi_{3D,1}} = & \frac{1}{3}(e^{i\frac{4\pi}{3}}\sqrt{2}\ket{2,0,0} + e^{i\frac{2\pi}{3}}\sqrt{2}\ket{0,2,0} +\sqrt{2}\ket{0,0,2} \\ - \ket{1,1,0} & + e^{i\frac{5\pi}{3}}\ket{1,0,1} + e^{i\frac{\pi}{3}}\ket{0,1,1}).
\end{aligned}
\end{equation}
From this equation one can see that for these computational basis inputs, no term completely vanishes from the state. 
In other words, the visibility in a bunching experiment using orthogonal output modes is limited to 50\%, as we also observed in the measurement shown in Fig. 3 in the main text. 

\begin{figure}[htb]
    \centering
    \includegraphics[width=\linewidth]{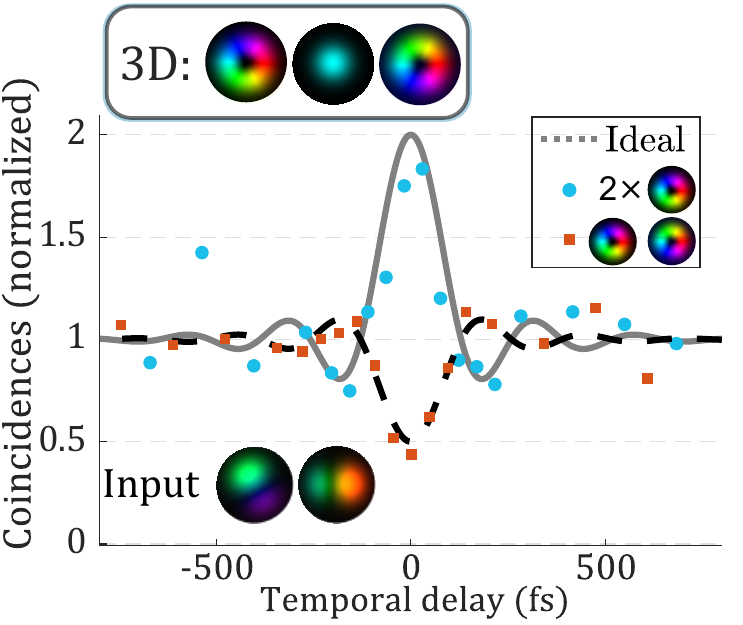}
    \caption{Interference measurements in a three-dimensional state space using superposition states as an input. 
    The results correspond well to the OAM-basis measurements shown in the main article. The error-bars for this data (and all the other data sets) are shown in Fig. \ref{fig:HD_rawdata} and the superposition states are described in equation\eqref{eq:3DMUB}.}
    \label{fig:3Dmub}
\end{figure}

One set of results that was omitted from the main text were measurements in the three-dimensional state space of another MUB (measurements shown in Fig. \ref{fig:3Dmub}). We used input modes defined by the states
\begin{equation}
    \label{eq:3DMUB}
    \begin{aligned}
        \ket{\Psi_{MUB2,2}} &= \frac{1}{\sqrt{3}}\left(e^{i\frac{2\pi}{3}}\ket{1}_{-1} + e^{i\frac{4\pi}{3}}\ket{1}_{0} +  \ket{1}_{+1}\right) \text{ and}\\
        \ket{\Psi_{MUB2,3}} &= \frac{1}{\sqrt{3}}\left(e^{i\frac{2\pi}{3}}\ket{1}_{-1} + \ket{1}_{0} + e^{i\frac{4\pi}{3}}\ket{1}_{+1}\right),
    \end{aligned}
\end{equation}
where the subscripts refer to the OAM value of the corresponding modes. Similar to the measurements done in the OAM-basis (shown in the main article), bi-photon interference leads to coalescence of photons with a visibility up to 50\% as can be seen in Fig. \ref{fig:3Dmub}.
When projecting the two photons onto orthogonal modes, i.e. $\hat{P}_{+1-1}$, we obtain a dip with a visibility of $54.1\%~\pm~4.5\%$, which is in very good agreement with the theoretically predicted $50\%$.
Similarly, we observe the bunching as a bump in the coincidence rate when projecting both photons on identical states, i.e. $\hat{P}_{+1+1}$, for which we obtain a visibility of $V = 78.4\%~\pm~11.7$.
Thus we see that, as with the 2D-transformation, photons in states of different 3D MUBs behave similarly to the computational basis modes.

\begin{figure}[htb]
    \centering
    \includegraphics[width=\linewidth]{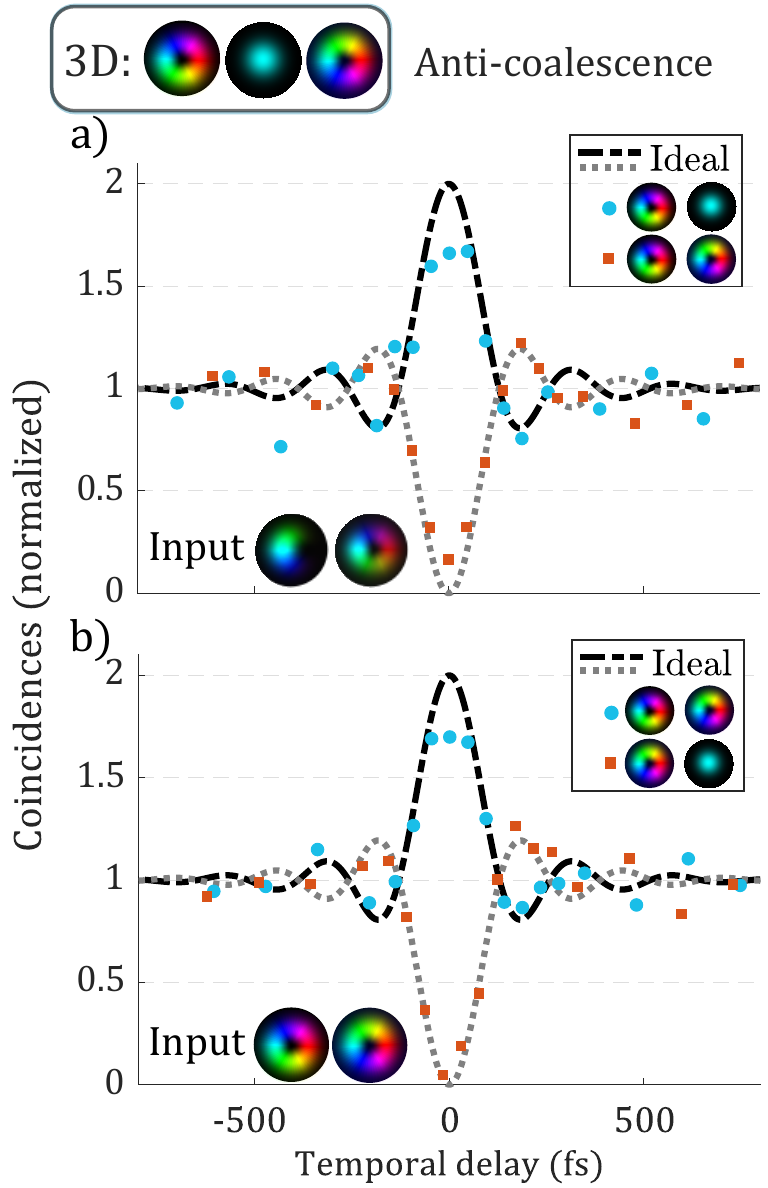}
    \caption{Anti-coalescence measurements in a three-dimensional basis. In a) we rotated our input states into an unbalanced superposition of the three OAM-basis states and achieved coalescence with one projection (orange squares) and anti-coalescence with a different projection (blue dots). In b) we moved the same unbalance from the input state into the unitary and achieved similar results. The measurement shown with blue dots in b) was also shown in the main article in Fig. 4a). The error-bars are again shown in Fig. \ref{fig:HD_rawdata} and the details on the input states and the unbalanced unitary can be found in equations (\ref{eq:unEQ_sup}-\ref{eq:RotUni}).}
    \label{fig:3DAnti}
\end{figure}

To observe anti-coalescence in the same $\hat{U}_3$ unitary, one can use the unbalanced superposition states 
\begin{equation}
\label{eq:unEQ_sup}
\begin{split}
    \ket{\Psi_{A1}} = \frac{1}{2\sqrt{3}} &[(\sqrt{2} + 1)\ket{1}_{-1} + (\sqrt{2} + 1)\ket{1}_{0} + \\ &(\sqrt{2} - 2)\ket{1}_{+1}]\\
    \ket{\Psi_{A2}} = \frac{1}{2\sqrt{3}} &[(\sqrt{2} - 1)\ket{1}_{-1} + (\sqrt{2} - 1)\ket{1}_{0} +\\ &(\sqrt{2} + 2)\ket{1}_{+1}],
\end{split}
\end{equation}
as an input.
The perfect coalescence and anti-coalescence that can be achieved with these input states was omitted from the main article but can be found in Fig. \ref{fig:3DAnti}a).
Here, it can also be seen that $\ket{\Psi_{A1}}$ is close to a two-mode superposition and  $\ket{\Psi_{A2}}$ is almost a pure $\ell = +1$ OAM-state. 

To observe anti-coalescence using pure OAM input states (Fig. 4a) of main text and Fig. \ref{fig:3DAnti}b)), we used the rotated unitary 
\begin{equation}
\label{eq:RotUni}
\begin{aligned}
    & \hat{U}_{Rot+3} = \hat{U}_3\hat{R}_3, \\ 
    & \hat{R}_3 =  
    \frac{1}{2\sqrt{3}}
    \begin{bmatrix}
    \sqrt{2} + 1 & \sqrt{2} - 1 & -2\sqrt{\frac{3}{2}}  \\
    \sqrt{2} - 2 & \sqrt{2} + 2 & 0 \\
    \sqrt{2} + 1 & \sqrt{2} - 1 & 2\sqrt{\frac{3}{2}}
    \end{bmatrix},
\end{aligned}
\end{equation}
where $\hat{R}_3$ is the rotation unitary derived from the superposition modes shown in equation (\ref{eq:unEQ_sup}).
As OAM modes are generated more efficiently compared to Gaussian modes, when using the amplitude and phase modulation technique, we adjusted the unitary in our experiment to be able to use OAM $\ell = \pm1$ states.
Mathematically this means that we change our basis state space to $\ket{\ell = -1,\ell = +1,\ell = 0}$ instead of $\ket{\ell = -1,\ell = 0,\ell = +1}$ when preparing the unitary.
While this change does not lead to any significant differences, it led to the observation of the anti-coalescence with a different projection measurement in Fig. \ref{fig:3DAnti}a) and Fig. \ref{fig:3DAnti}b).
The obtained output state using, $\ket{\ell = -1,\ell = 0,\ell = +1}$, is 
\begin{equation}
\begin{split}
    \ket{\Psi_{Rot+3}} = &\frac{\sqrt{2}}{4}e^{-i\pi/3}\ket{2,0,0} + \frac{\sqrt{2}}{2}\ket{0,2,0} \\+ &\frac{\sqrt{2}}{4}e^{i\pi/3}\ket{0,0,2} + \frac{1}{2}e^{i\pi}\ket{1,0,1}.
\end{split}
\end{equation}
Thus, the probability of the $\ket{1,0,1}$ state is $0.25$ after interference whereas it is $0.125$ without interference, resulting in the observed coalescence.
This is then observed as a HOM-like bump shown in Fig. \ref{fig:3DAnti}b).
The $\ket{1,1,0}$ and $\ket{0,1,1}$ states have disappeared during the bunching before which they both had the probability $0.25$.
This results in a HOM-like dip, shown for $\ket{0,1,1}$ in Fig. \ref{fig:3DAnti}b).
The anti-coalescence bump in Fig. \ref{fig:3DAnti}a), projection $\hat{P}_{0+1}$, had a visibility of $71.7\%~\pm~7.9\%$ and the corresponding dip between the other mode pair, $\hat{P}_{+1-1}$, had a visibility of $84.2\%~\pm~4.5\%$.
For the unbalanced unitary the accompanying coalescence dip shown in Fig. \ref{fig:3DAnti}b) had a visibility of $94.9\%~\pm~4.0\%$.
The observed anti-coalescence is a result of our ability to tune the relative phases between different output channels, by rotating the input states into any arbitrary superposition.
The same type of anti-coalescence can also be achieved for a lossy beamsplitter \cite{Uppu:16}.

\begin{figure*}
\centering
    \includegraphics[width=0.9\linewidth]{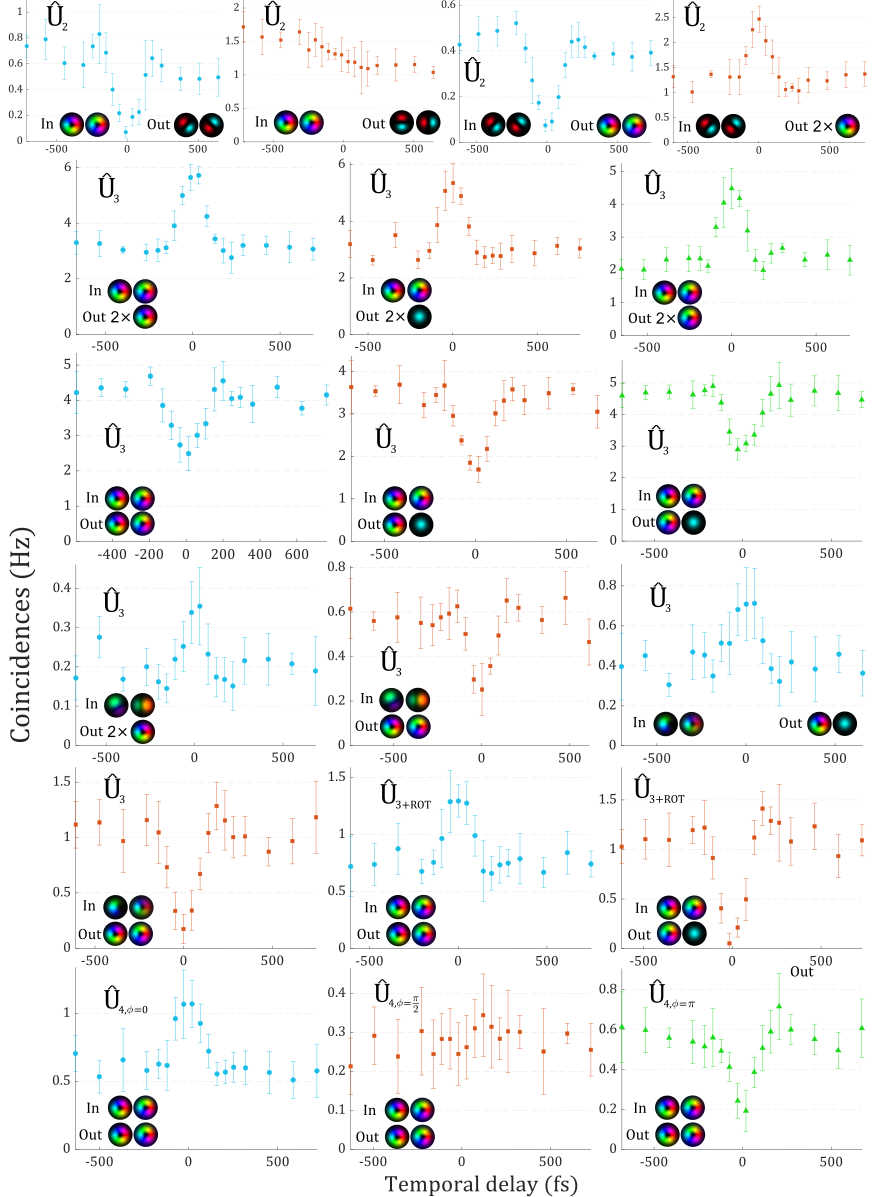}
    \caption{Data with errorbars from the three and four-dimensional measurements. 
            The plots are shown in the same order that they appear in the main article. 
            The corresponding unitary, input modes, and output modes are inset into the plots.
            The standard errors were calculated from multiple consecutive measurements, the number of measurements done per data point (starting from top left) were 
            5, 5, 5, 5,
            5, 6, 6,
            5, 5, 5,
            5, 6, 10,
            10, 10, 10,
            10, 6, and 6, in each plot respectively.}
    \label{fig:HD_rawdata}
\end{figure*}

%\section{Four-dimensional unitary}
%In four dimensions, our unitary of choice was 
%\begin{equation}
%    \hat{U}_4 = \frac{1}{2}
%    \begin{bmatrix}
%    1 & 1 & 1 & 1 \\
%    1 & e^{i\varphi} & -1 & -e^{i\varphi}  \\
%    1 & -1 & 1 & -1 \\
%    1 & -e^{i\varphi} & -1 & e^{i\varphi}
%    \end{bmatrix},
%    \label{eq:U4}
%\end{equation}
%where $\varphi$ corresponds to the free adjustable phase that can range between $[0,2\pi]$ %\cite{mattle1995nonclassical}.

\bibliography{Supple.bib}